\documentclass[aps,prb,reprint,showpacs,twocolumn,groupedaddress]{revtex4-1}

\usepackage{graphicx}

\newcommand*{\Ham}{{\cal{H}}}
\newcommand*{\HamB}{{\cal{H}}_B}
\newcommand*{\HamT}{{\cal{H}}_T}

\newcommand*{\HamL}{{\cal{H}}_{leads}}
\newcommand*{\Heff}{\mbox{\boldmath$H$}}
\newcommand*{\SMM}{\mbox{\boldmath$S$}}

\newcommand*{\svec}{\mbox{\boldmath$\sigma$}}
\newcommand*{\sx}{\sigma_x}
\newcommand*{\sy}{\sigma_y}
\newcommand*{\sz}{\sigma_z}
\newcommand*{\torque}{\mbox{\boldmath$\tau$}}

\newcommand*{\ex}{\mbox{\boldmath$e$}_x}
\newcommand*{\ey}{\mbox{\boldmath$e$}_y}
\newcommand*{\ez}{\mbox{\boldmath$e$}_z}
\newcommand*{\eS}{\mbox{\boldmath$e$}_S}

\newcommand*{\dvec}{\mbox{\boldmath$d$}}
\newcommand*{\gvec}{\mbox{\boldmath$g$}}
\newcommand*{\fvec}{\mbox{\boldmath$f$}}
\newcommand*{\jvec}{\mbox{\boldmath$j$}}
\newcommand*{\nvec}{\mbox{\boldmath$n$}}

\newcommand*{\vF}{\mbox{\boldmath$v$}_F}

\begin{document}


\title{Spin-precession-assisted supercurrent in a superconducting quantum point contact coupled to a single-molecule magnet}


\author{C. Holmqvist}

\affiliation{Fachbereich Physik, Universit\"at Konstanz, D-78457 Konstanz, Germany}

\author{W. Belzig}

\affiliation{Fachbereich Physik, Universit\"at Konstanz, D-78457 Konstanz, Germany}

\author{M. Fogelstr\"om}

\affiliation{Department of Microtechnology and Nanoscience - MC2, Chalmers University of Technology,
SE-412 96 G\"oteborg, Sweden}



\date{\today}

\begin{abstract}
The supercurrent of a quantum point contact coupled to a nanomagnet strongly depends on the dynamics of the nanomagnet's spin. We employ a fully microscopic model to calculate the transport properties of a junction coupled to a spin whose dynamics is modeled as Larmor precession brought about by an external magnetic field and find that the dynamics affects the charge and spin currents by inducing transitions between the continuum states below the superconducting gap edge and the Andreev levels. This redistribution of the quasiparticles leads to a non-equilibrium population of the Andreev levels and an enhancement of the supercurrent which is visible as a modified current-phase relation as well as a non-monotonous critical current as function of temperature. The non-monotonous behavior is accompanied by a corresponding change in spin-transfer torques acting on the precessing spin and leads to the possibility of using temperature as a means to tune the back-action on the spin.
\end{abstract}


\maketitle


\section{Introduction}
Spintronics devices, which utilize the spin degree of freedom, have already revolutionized read-out technology used in hard drives. \cite{baibich1988,binasch1989} In conventional spintronics, the transport properties of a device typically depend on the relative orientation of the spins of electrons with respect to a reference, which may be a magnetic field or a magnetization direction of a ferromagnetic layer. \cite{wolf2001} The challenge of downsizing electronic devices has lead to the study of transport properties of non-magnetic single-molecule devices such as diodes \cite{elbing2005} and transistors \cite{park2002,osorio2008,yu2005} as well as devices containing single-molecule magnets (SMMs). \cite{bogani2008} The interest in SMMs stems from their long relaxation times at low temperatures \cite{christou2000} and their display of a wide range of quantum physics phenomena. \cite{christou2000,wernsdorfer1999} Studies on SMM devices include for instance three-terminal devices, \cite{heersche2006,jo2006,henderson2007,grose2008,zyazin2010,roch2011,haque2011} supramolecular spin valves, \cite{urdampilleta2011} and inelastic tunneling spectroscopy. \cite{kahle2012}

\begin{figure*}[t]
\includegraphics[width=1.90\columnwidth,angle=0]{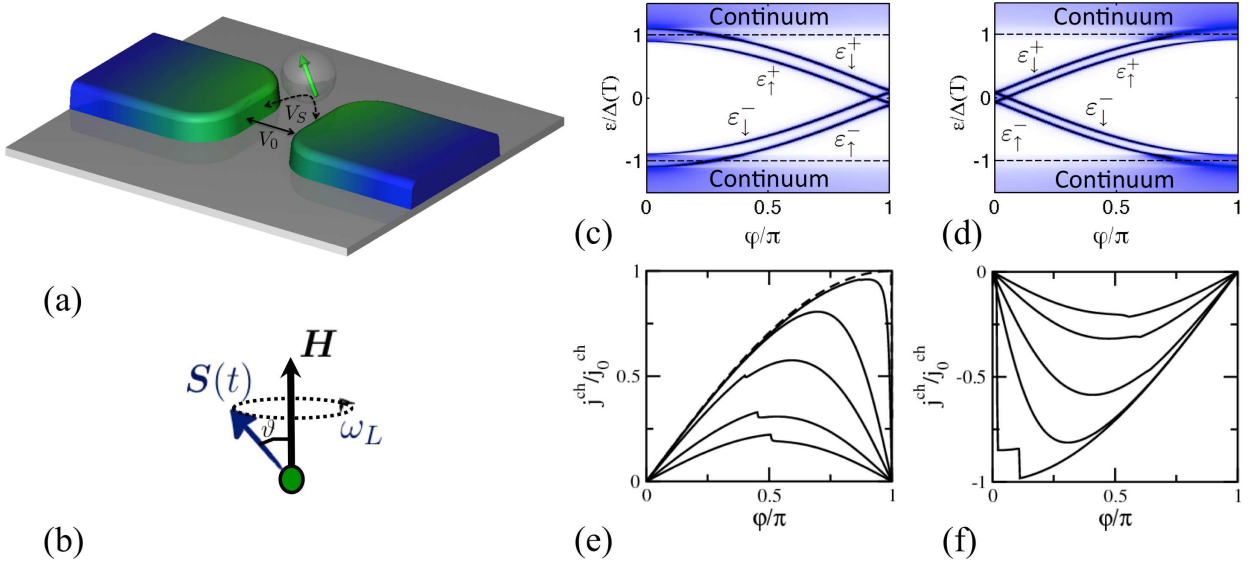}
\caption{ 
(Color online) (a) Two superconducting leads are coupled over the spin of a nanomagnet. The hopping amplitude for spin-independent tunneling is given by $V_0$, while $V_S$ is the coupling between the tunneling electrons and the spin of the nanomagnet. (b) The spin of the nanomagnet, $\SMM$, precesses with the Larmor frequency, $\omega_L$, due to an external effective magnetic field, $\Heff$, that is applied at an angle $\vartheta$ with respect to the orientation of the spin. The density of states in a rotating frame (see definition in the text) is plotted for a transparent junction ($v_{T}=1$) with hopping strengths (c) $v_0=v_{T}\cos(0.1\pi/2)\approx 0.9877, v_S=v_{T}\sin(0.1\pi/2)\approx 0.1564$ and (d) $v_0=v_{T}\cos(\pi/2)=0, v_S=v_{T}\sin(\pi/2)=1$ at temperature $T/T_c\rightarrow 0$. The Andreev levels, which can be seen as sharp states inside the gap, have here for visibility reasons been given an artificial broadening, $0^+$, and the continuum density of states is normalized to 1 away from the gap edge. The upper (lower) Andreev levels, $\varepsilon^+$ ($\varepsilon^-$), are given an effective Zeemann splitting ($\varepsilon^{+(-)}\rightarrow\varepsilon^{+(-)}_{\uparrow,\downarrow}$) by the spin precession. The spin precession couples the Andreev levels and the continuum states as well as the states $\varepsilon_\uparrow^{+(-)}$ and $\varepsilon_\downarrow^{+(-)}$. (e) The current-phase relation (CPR) is plotted for $v_0=\cos(0.1\pi/2), v_S=\sin(0.1\pi/2)$ at temperatures (solid lines from top to bottom) $T/T_c=7.5\cdot 10^{-4}$, $0.25$, $0.50$, $0.75$ and $0.85$. The dashed line shows the CPR for $(v_0=1, v_S=0)$ at temperature $T/T_c\rightarrow 0$. (f) The CPR for a junction with $(v_0=0, v_S=1)$ at temperatures (from bottom to top) $T/T_c=7.5\cdot 10^{-4}$, $0.25$, $0.50$, $0.75$ and $0.85$. In both (e) and (f), the CPRs develop a peak as the temperature increases. Figures (c) - (f) are calculated for precession angle $\vartheta=\pi/4$ and frequency $\omega_L=0.2\Delta$.
}
\label{fig:junction}
\vspace*{-0.3truecm}
\end{figure*}

Interesting spin phenomena may occur when ferromagnets are combined with superconductors (see \onlinecite{eschrig2011} and references therein). Cooper pairs in a conventional superconductor have spin-singlet pairing which, if the superconductor is interfaced with a ferromagnet, extend into the ferromagnet. However, the exchange field inside the ferromagnet tries to align the two spins of the Cooper pairs and hence breaks the Cooper pairs apart resulting in a rapid decay of the superconducting correlations inside the ferromagnet. For the same reasons, the critical current of a Josephson junction with a ferromagnetic layer sandwiched between the two superconductors decays rapidly with increasing thickness of the ferromagnetic layer. \cite{bulaevskii1977,ryazanov2001,kontos2002,buzdin2005} On the other hand, if weakly ferromagnetic interfaces with magnetization directions differing from the magnetization direction of the ferromagnetic layer are inserted, the spin-singlet correlations may be transformed into spin-triplet correlations which can survive over a long range within the ferromagnet layer. \cite{bergeret2001,bergeret2005,houzet2007,braude2007,eschrig2008} As a result of this non-collinear magnetization of the ferromagnetic layer, the critical current decays similarly to a supercurrent in a non-magnetic metal with increasing junction length. \cite{keizer2006,khaire2010} Also interaction between spin and charge supercurrents has attracted attention. \cite{waintal2001,waintal2002,zhao2008,shomali2011} Recently, interest in coupling between the dynamics of magnetic moments and Josephson currents has increased. \cite{braude2008,michelsen2008,petkovic2009,barnes2011,mai2011,linder2011}

In Refs. \onlinecite{holmqvist2011} and  \onlinecite{teber2010}, it was found that spin-singlet to spin-triplet conversion can be generated by a nanomagnet such as a SMM or a ferromagnetic nanoparticle coupled to a Josephson junction \cite{kazumov2005} provided that the magnetization direction of the nanomagnet precesses. \cite{houzet2008} The spin-triplet correlations enable spin currents to exist close to the junction interface despite the s-wave nature of the superconducting leads. The spin currents generate a spin-transfer torque acting on the nanomagnet and its effect may be measurable in a ferromagnetic resonance (FMR) experiment \cite{bell2008} as was suggested in Ref. \onlinecite{holmqvist2011}. Refs. \onlinecite{holmqvist2011} and  \onlinecite{teber2010} focused on charge and spin currents as well as spin-triplet correlations at zero temperature and in equilibrium junctions. Here, we instead investigate the effects of the spin precession on the critical current and find that the critical current is enhanced at high temperatures. This enhancement is due to a redistribution of the population of states caused by the precessing spin and is also visible in the current-phase relation (CPR). The redistribution of quasiparticles also affects the spin currents, which change abruptly as a function of temperature when the critical current is enhanced.

The paper is organized as follows: In section \ref{sec:model}, the model of the superconducting point contact containing the molecular magnet is described. The approach to solving the spin-active boundary condition created by the molecular magnet is then described in section \ref{sec:approach}. Subsequently, the results are discussed in section \ref{sec:results} starting with the Andreev levels and their population. Then, the charge currents and spin currents are discussed along with the induced spin-triplet correlations associated with the spin currents. Finally, the results are summarized in section \ref{sec: conclusions}. Appendix \ref{sec: appendix} details some of the calculations in section \ref{sec:results}.

\section{Model}\label{sec:model}
Here, we consider the same junction model as in Refs. \onlinecite{holmqvist2011} and \onlinecite{teber2010}, i.e. two superconducting leads coupled over a nanomagnet, schematically depicted in Fig. \ref{fig:junction}(a). The nanomagnet may be a SMM or a ferromagnetic nanoparticle and is subjected to an external effective magnetic field, $\Heff$, which couples to the spin of the nanomagnet, $\SMM$, through the Hamiltonian $\HamB=-\gamma\Heff\cdot\SMM$, where $\gamma$ is the gyromagnetic ratio (see Fig. \ref{fig:junction}(b)). We treat the spin classically and assume that the effective magnetic field is applied at an angle $\vartheta$ with respect to the spin. In this classical limit, the effective magnetic field results in a torque acting on the spin, $\dot{\SMM}=-\gamma\Heff\times\SMM$, causing the spin to precess with the Larmor frequency, $\omega_L=\gamma |\Heff |$. The effective magnetic field is taken to be applied along the $z$ axis and hence the spin, $\SMM$, can be parameterized as $\SMM(t)=S\eS(t)$ where $S=|\SMM|$ is the magnitude of the spin and $\eS(t)=\cos(\omega_Lt)\sin\vartheta\ex+\sin(\omega_Lt)\sin\vartheta\ey+\cos\vartheta\ez$.
The two superconducting leads are then coupled over the nanomagnet leading to a complete Hamiltonian given by $\Ham=\HamL+\HamB+\HamT$, \cite{zhu2003,zhu2004} where $\HamL$ describing the leads is the BCS Hamiltonian with the superconducting order parameter $\Delta(T)e^{\pm i\varphi/2}$, where "$+$" ("$-$") refers the right (left) lead.
$\HamT$ is the tunneling Hamiltonian,
$\HamT= \sum_{k \sigma;k^\prime\sigma^\prime} c^\dagger_{L,k\sigma} V_{k\sigma;k^\prime\sigma^\prime} c_{R,k^\prime\sigma^\prime} +h.c.$,
where the hopping amplitude is given by $V_{k\sigma;k^\prime \sigma^{\prime}}=(V_0 \delta_{\sigma \sigma^{\prime}}+V_S(\SMM (t) \cdot \svec)_{\sigma \sigma^{\prime}})\delta(k-k^{\prime})$ and $\svec=(\sx,\sy,\sz)$ consists of the Pauli spin matrices.
The first term of the hopping amplitude describes spin-independent tunneling, while the second term corresponds to spin-dependent tunneling and depends on the instantaneous direction of the spin, $\SMM$. The projection of $\SMM$ along the $z$ axis leads to different tunnel probabilities for spin-up and -down quasiparticles while the transverse components of the spin lead to spin flips.
Moreover, we assume that the tunneling quasiparticles are affected only by the exchange interaction with the nanomagnet and that the external magnetic field does not couple to the leads or directly to the tunneling quasiparticles.

\section{Approach}\label{sec:approach}
We use non-equilibrium Green's functions in the quasiclassical approximation \cite{eilenberger1968,larkin1968,eliashberg1971} to calculate the current through the junction. \cite{cuevas2001,kopu2004} The quasiclassical Green's functions are propagators for quasiparticle on classical trajectories. An incoming propagator describes quasiparticles with trajectories leading into the junction interface, i.e. $\vF\cdot\hat{\nvec}<0$ where $\hat{\nvec}$ is the surface normal and $\vF$ is the group velocity of the electron-like quasiparticles. Conversely, outgoing propagators describe quasiparticles having trajectories leading out from the interface, $\vF\cdot\hat{\nvec}>0$.
The spin-singlet superconductivity of the uncoupled leads can be described by spin-scalar Green's functions.
However, to accommodate for the effects of the spin-dependent scattering caused by the nanomagnet, the Green's functions are parametrized as consisting of both spin-scalar ($s$) and spin-vector ($t$) components as \cite{serene1983}
\begin{equation}\label{scalar and triplet green's function}
\hat{g}^{X}=\left(\begin{array}{cc}g^X_s+ \gvec^X_t \cdot \svec & (f^X_s+\fvec^X_t\cdot\svec)i\sy \\ 
i\sy ({\tilde f}^X_s + {\tilde{\fvec}}^X_t \cdot \svec)& {\tilde g}^X_s - \sigma_y ({\tilde{\gvec}}^X_t\cdot \svec )\sigma_y 
\end{array}\right),
\end{equation}
where $X$ stands for retarded ($R$), advanced ($A$) and Keldysh ($K$) Green's functions and the hat ("$\,\hat{\,}\,$") denotes matrices in Nambu-spin space. The components $f^X_s$ and $\fvec^X_t$ are the spin-singlet and spin-triplet components of the anomalous Green's functions, respectively (similarly for $g^X_s$ and $\gvec^X_t$).

The precessing magnetization of the nanomagnet constitutes a time-dependent, spin-active boundary condition that can be solved by applying the unitary transformation
\begin{equation}
\hat {\cal {U}}(t)=\left(\begin{array}{cc}{\rm e}^{-i \frac{\omega_L}{2}t \sz} &0 \\0& {\rm e}^{i \frac{\omega_L}{2}t \sz} \end{array}\right),
\label{utrans}
\end{equation}
which results in a transformation to a rotating frame of reference where the Fermi surfaces of the spin-up and -down bands are shifted with $\mp\omega_L/2$. This transformation also shifts the gap edges as follows: The upper gap edge, $\Delta^+(T)=\Delta(T)$, is shifted as $\Delta^+(T)\rightarrow \Delta^+_{\uparrow,\downarrow}(T)$, where $\Delta^+_{\sigma}(T)=\Delta(T)-\sigma \omega_L/2$ and $\sigma=1 (-1)$ for spin $\uparrow(\downarrow)$. The lower gap edge, $\Delta^-(T)=-\Delta(T)$, is correspondingly modified as $\Delta^-(T)\rightarrow \Delta^-_{\uparrow,\downarrow}(T)$, where $\Delta^-_{\sigma}(T)=-\Delta(T)-\sigma \omega_L/2$.
In this rotating frame, the precessing spin, $\SMM(t)$, now appears static and points along the direction $\eS=\cos\vartheta \ez+\sin\vartheta \ex$. Replacing the hopping amplitudes $V_0$ and $V_S$ by their Fermi surface limits, $v_0=\pi N_F V_0$ and $v_S=\pi N_F S V_S$ where $N_F$ is the normal density of states at the Fermi energy, the hopping element in Nambu-spin space reads
\begin{equation}
\hat {v}=\left(\begin{array}{cc} v_0 + v_S \eS \cdot \svec &0 \\0& v_0 -\sy (v_S \eS \cdot \svec) \sy \end{array}\right)
\label{v-hat}
\end{equation}
and $\check{v}=\hat{v}\check{1}$ in Keldysh space.

As a first step towards solving the boundary condition, the interface is treated as an impenetrable surface and the quasiclassical Green's functions are found by solving the Eilenberger equation \cite{eilenberger1968,larkin1968,eliashberg1971} in each lead separately. The resulting solutions, denoted by $\check{g}^0_\alpha$ with $\alpha=L,R$ indicating the left or right lead, are then connected across the interface using a quasiclassical t-matrix equation, \cite{caroli1971,buchholtz1979,martinrodero1994}
\begin{equation}\label{eq: t matrix}
\check{t}_\alpha(\varepsilon)=\check{\Gamma}_\alpha(\varepsilon)+\check{\Gamma}_\alpha(\varepsilon) \check{g}^0_\alpha(\varepsilon) \check{t}_\alpha(\varepsilon),
\end{equation}
which takes the hopping Hamiltonian $\HamT$ into account through
\begin{equation}
\check{\Gamma}_{L/R}(\varepsilon)= \check{v} \check{g}^0_{R/L}(\varepsilon) \check{v}.
\end{equation} 
In this way, the full propagators for the junction can be obtained as
\begin{equation}\label{eq: g in out}
\check{g}^{i,o}_\alpha(\varepsilon)= \check{g}^{0}_\alpha(\varepsilon)+(\check{g}^{0}_\alpha(\varepsilon)\pm i \pi \check{1}) \check{t}_\alpha(\varepsilon) (\check{g}^{0}_\alpha(\varepsilon)\mp i \pi \check{1}),
\end{equation}
where $\check{g}^{i(o)}$ denotes the incoming (outgoing) propagator.
The difference between the incoming and the outgoing propagators then give expressions for the charge Josephson current and the spin Josephson current per conduction channel as
\begin{eqnarray}
j^{ch}_\alpha(t) &=& \frac{e}{2\hbar} \int \frac{d \varepsilon}{8 \pi i} \mbox{Tr} [ \hat{\tau}_3 
(\hat{g}^{i,<}_{\alpha}(\varepsilon,t)  - \hat{g}^{o,<}_{\alpha}(\varepsilon,t)) ]; \label{chargecurr} \\
\jvec^{s}_\alpha(t) &=&  \frac{1}{4}  \int \frac{d \varepsilon}{8 \pi i}  \mbox{Tr} [ \hat{\tau}_3 \hat{\svec} 
( \hat{g}^{i,<}_{\alpha}(\varepsilon,t) - \hat{g}^{o,<}_{\alpha}(\varepsilon,t)) ]
\label{spincurr}
\end{eqnarray}
where $\hat \tau_3={\rm diag}(1,-1)$ and $\hat{\svec}={\rm diag}(\svec,-\sy\svec\sy)$. The lesser ("$<$") Green's functions is obtained as $\hat{g}^<=\frac{1}{2}(\hat{g}^{K}-\hat{g}^{R}+\hat{g}^{A})$. The details are described in Refs. \onlinecite{holmqvist2011} and \onlinecite{teber2010}.

\section{Results}\label{sec:results}
Scattering processes between two superconductors may lead to constructive interference and the appearance of Andreev levels. \cite{pillet2010} In the presence of a precessing spin, a tunneling quasiparticle may gain (lose) energy $\omega_L$ while simultaneously flipping its spin from down (up) to up (down). These additional tunneling processes lead to a modified Andreev level spectrum whose details depend on a number of parameters: The ratio $v_0/v_S$ determines whether the junction is in a $0$ or a $\pi$ state, depending on if $v_0/v_S>1$ or $v_0/v_S<1$, respectively. \cite{holmqvist2011,teber2010}
The Larmor frequency, $\omega_L$, determines the amount of energy a quasiparticle may gain or lose during tunneling across the junction. The precession angle, $\vartheta$, determines the amount of scattering between the spin-up and -down bands. The population of the Andreev states is modified by the temperature, $T$, but also by the scattering processes generated by the precessing spin. This modification of the Andreev level population at finite temperature is the focus of this paper.

\subsection{Andreev levels}\label{subsec: andreev levels}

In the case of a static spin, for which one can take $\omega_L=0$ and $\vartheta=0$, the Andreev levels are given by
\begin{equation}\label{AL theta=0}
\varepsilon^\pm=\pm\Delta\sqrt{1-{\cal D}_0\sin^2\frac{\varphi}{2}-{\cal D}_S\cos^2\frac{\varphi}{2}}
\end{equation}
where
\begin{equation}
{\cal D}_{0(S)}=\frac{4v_{0(S)}^2}{1+2(v_0^2+v_S^2)+(v_0^2-v_S^2)^2}
\end{equation}
are transmission probabilities determined by the hopping amplitudes $v_0$ and $v_S$. The two signs of $\varepsilon^\pm$ describe the two Andreev level branches; one that exists below the Fermi surface, $\varepsilon^-$, and one that exists above the Fermi surface, $\varepsilon^+$. From Eq. (\ref{AL theta=0}), it is clear that the junction is in the $0$ state if $v_0>v_S$, and in the $\pi$ state if $v_0<v_S$. \cite{bulaevskii1977}

In the limit of $v_S\rightarrow 0$, but with an arbitrary precession frequency, the Andreev levels in the rotating frame display an effective Zeemann splitting, $\varepsilon^{+(-)}\rightarrow \varepsilon^{+(-)}_{\uparrow,\downarrow}$. The two branches are then given by
\begin{eqnarray}
\varepsilon_{\sigma}^+&=&\Delta \bigg\{ \sqrt{1-{\cal D}_0 \sin^2\left(\frac{\varphi}{2}\right) } -\sigma \frac{\omega_L}{2\Delta} \bigg\} \nonumber \\ 
\varepsilon_{\sigma}^-&=&-\Delta \bigg\{ \sqrt{1-{\cal D}_0 \sin^2\left(\frac{\varphi}{2}\right) } +\sigma \frac{\omega_L}{2\Delta} \bigg\} .
\label{eq: AL split}
\end{eqnarray}
Note that these states are eigenstates only when there is no spin-flip scattering and that spin-flip scattering leads to scattering between $\varepsilon^{+(-)}_\uparrow$ and $\varepsilon^{+(-)}_\downarrow$.
The Andreev levels are visible in the density of states as sharp subgap states. The time-averaged density of states is evaluated as
\begin{equation}
\rho_\alpha(\varepsilon,\varphi)=-\frac{1}{8 \pi} \Im \{ \mbox{Tr} [ \hat{\tau}_3 
(\hat{g}^{d,i,R}_{\alpha}(\varepsilon,\varphi)+ \hat{g}^{d,o,R}_{\alpha}(\varepsilon,\varphi)) ] \},
\end{equation}
where $\hat{g}^{d,i/o,R}_{\alpha}(\varepsilon,\varphi)$ is the diagonal component of $\hat{g}^{i/o,R}_{\alpha}(\varepsilon,\varphi,t)$ and is time independent (see Ref. \onlinecite{holmqvist2011}). 
In Fig. \ref{fig:junction}(c), the density of states is plotted in the limit $v_S\ll v_0$. The sharp states inside the superconducting gap are the Andreev levels, which are split into $\varepsilon^\pm_\uparrow$ and $\varepsilon^\pm_\downarrow$. These Andreev levels are well described by Eq. (\ref{eq: AL split}) and their splitting is given by $\omega_L$.

Increasing the spin-dependent hopping amplitude, $v_S$, results in scattering of quasiparticles between the split Andreev levels belonging to the same branch. In general, the Andreev levels can be described by
\begin{equation}
\epsilon_{\sigma}^{\pm}=\pm\Delta\sqrt{1+\Phi(v_0,v_S,\omega_L,\vartheta,\varphi) +\Xi_{\sigma}(v_0,v_S,\omega_L,\vartheta,\varphi) },
\end{equation}
where an analytical expression for $\Phi$ can be found as
\begin{eqnarray}
\Phi(v_0,v_S,\omega_L,\vartheta,\varphi)&=&-{\cal D}_0(\vartheta) \sin^2\left(\frac{\varphi}{2}\right) \\ \nonumber
&\quad& -{\cal D}_S(\vartheta) \cos^2\left(\frac{\varphi}{2}\right)+ \left(\frac{\omega_L}{2\Delta}\right)^2
\end{eqnarray}
and the transmission probabilities, ${\cal D}_S$ and ${\cal D}_0$, depend on the precession angle $\vartheta$. Defining $\delta=1+2(v_0^2+v_S^2)+(v_0^2-v_S^2)^2$, the transmission probabilities are given by
\begin{equation}
{\cal D}_S(\vartheta)=2v_S^2\left( \frac{1}{\delta} + \frac{\cos{(2\vartheta)}}{\delta-4v_S^2\sin^2\vartheta}+\frac{2v_S^2\sin^2(2\vartheta)}{(\delta-4v_S^2\sin^2\vartheta)^2}\right)
\end{equation}
and
\begin{equation}
{\cal D}_0(\vartheta)=2v_S^0\left( \frac{1}{\delta} + \frac{\cos{(2\vartheta)}}{\delta-4v_S^2\sin^2\vartheta}\right).
\end{equation}
The function $\Xi_{\sigma}(v_0,v_S,\omega_L,\vartheta,\varphi)$ provides the Zeemann splitting which in the general case also depends on $\vartheta$. 
An analytical expression for $\Xi_{\sigma}$ can in principle be obtained, but is too involved to allow for simple analytical analysis. The effects of this term are, however, numerically analyzed below.
The density of states for a junction with ($v_0=0,v_S=1$) is shown in Fig. \ref{fig:junction}(d). The Zeemann splitting in this case, $v_S\gg v_0$, can be approximated with $\sim \omega_L\cos\vartheta$.

\subsection{Enhancement of the critical current}\label{subsec: enhancement of the critical current}

Each Andreev level carries a certain amount of current that is weighted by the Andreev level occupation.
In equilibrium for a static spin, the amount of current each Andreev level, $\varepsilon^{\pm}$, carries is $(2e/\hbar)\partial \varepsilon^{\pm}/\partial\varphi$ while the population of the quasiparticle states is given by the Fermi distribution function, $\phi_0$. \cite{shumeiko1997,beenakker1991} In this equilibrium situation, the lesser Green's functions in Eq. (\ref{chargecurr}) can be written as $g^{i/o,<}_{\alpha,\sigma\sigma}(\varepsilon,\varphi)=-2\pi\rho^{i/o}_{\alpha,\sigma\sigma}(\varepsilon,\varphi)\phi^<_0(\varepsilon)$ using the partial density of states $\rho^{i/o}_{\alpha,\sigma\sigma'}(\varepsilon,\varphi)=\left(-\frac{1}{8\pi} \right) \Im \{ (\hat{g}^{i/o,R}_{\alpha})_{\sigma\sigma'}(\varepsilon,\varphi) \}$. 
Hence, in equilibrium, the charge current is given by
\begin{equation}
j^{ch}(\varphi)=\frac{e}{\hbar} \frac{({\cal D}_0-{\cal D}_S) \Delta\sin(\varphi)  \tanh[\varepsilon^+(\varphi)/2T]  }{\sqrt{1-{\cal D}_0\sin^2(\varphi/2)-{\cal D}_S\cos^2(\varphi/2)}}.
\end{equation}
 
\begin{figure}[!htb]
\includegraphics[width=0.97\columnwidth,angle=0]{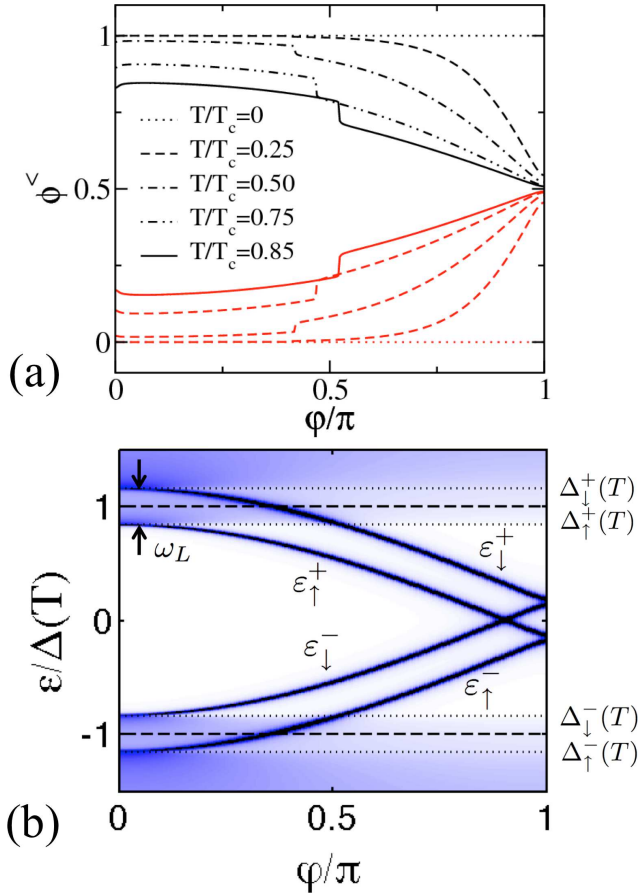}
\caption{\label{fig:population}
(Color online) (a) Population of the lower Andreev level $\varepsilon_\uparrow^-$ (black lines) and the upper Andreev level $\varepsilon_\downarrow^+$ (red lines) as a function of phase difference, $\varphi$, for a $0$ junction with hopping amplitudes $v_0=\cos(0.1\pi/2)$ and $v_S=\sin(0.1\pi/2)$, precession frequency $\omega_L=0.2\Delta(T=0)$ and tilt angle $\vartheta=\pi/4$. At temperature $T/T_c=7.5\cdot10^{-4}$ (dotted lines), the population of the Andreev levels is unaffected by the phase difference. At higher temperatures, $T/T_c=0.25$, $0.50$, $0.75$ and $0.85$ (solid lines), the population changes abruptly at a phase difference $\varphi_p$ leading to a jump in the CPRs shown in Fig. \ref{fig:junction}. (b) Density of states for the junction in (a) at temperature $T/T_c=0.85$. The dashed lines mark $\Delta^\pm(T)=\pm\Delta(T)$. The dotted lines denote $\Delta^\pm_\sigma(T)=\Delta^\pm(T)-\sigma\omega_L/2$. The population step in panel (a) occurs at the phase difference $\varphi_p$ given by the condition $\omega_L=|\Delta^\pm_\sigma(T)|-|\varepsilon^\pm_\sigma(T,\varphi)|$.}
\end{figure}

In a non-equilibrium situation where a superconducting point contact contains a spin precessing with a finite frequency, the current-phase relation is modified. As was shown in Refs. \onlinecite{holmqvist2011} and \onlinecite{teber2010}, the Josephson charge current is time independent and in the case of dominating spin-dependent tunneling and zero temperature, the CPR exhibits abrupt jumps as a function the superconducting phase difference, $\varphi$. As the temperature is increased, the abrupt jumps are smoothed out and a new step at a phase difference $\varphi_p$ develops as can be seen in Fig. \ref{fig:junction}(e) and (f). This step, which consequently gives a peak in the CPR at $\varphi_p$, is the result of an enhanced Josephson current at high temperatures for phase differences in the interval $0\le\varphi\le\varphi_p$ for $0$ junctions and for $\varphi_p\le \varphi\le \pi$ for $\pi$ junctions. To understand this current enhancement, we now turn to the population of the Andreev levels.

The population of the spin band $\sigma$ in a non-equilibrium situation can analogously to the equilibrium case be defined as
\begin{equation}
\phi^{i/o,<}_{\sigma}(\varepsilon,\varphi)= g^{i/o,<}_{0,\sigma\sigma}(\varepsilon,\varphi)/[-2\pi\rho^{i/o}_{\sigma\sigma}(\varepsilon,\varphi)].
\end{equation}
In Fig. \ref{fig:population}(a), the occupation of the Andreev levels $\varepsilon_\uparrow^-$ (black lines) and $\varepsilon_\downarrow^+$ (red lines) in a junction with $(v_0=\cos(0.1\pi/2),v_S=\sin(0.1\pi/2))$ is plotted as a function of phase difference, $\varphi$, for temperatures $T/T_c=7.5\cdot10^{-4}$ (dotted lines), $0.25$, $0.50$, $0.75$ and $0.85$ (solid lines). At temperature $T\rightarrow 0$, the lower Andreev level, $\varepsilon_\uparrow^-$, is fully occupied while the upper Andreev level, $\varepsilon_\downarrow^+$, is unoccupied. At higher temperatures, the lower Andreev level is not fully occupied while the upper correspondingly has a finite occupation. In addition, there is also an abrupt change in the population corresponding to the jump in the CPR at phase difference $\varphi_p$. The abrupt change in the population is an effect of the spin-flip scattering processes in which a quasiparticle interacting with the precessing spin may gain or lose energy $\omega_L$. These scattering processes couple the Andreev levels with the continuum states provided that $\omega_L\ge |\Delta^{+(-)}_{\sigma} (T)|-|\varepsilon_{\sigma}^{+(-)}|$, see Fig. \ref{fig:population}(b). The coupling causes quasiparticles to be promoted to the lower Andreev level from the continuum below the gap edge. Quasiparticles in the upper Andreev level are similarly scattered into the upper continuum. If the lower (upper) Andreev level is not completely filled (unoccupied), which is the case at finite temperature, the coupling leads to a repopulation (emptying) of the Andreev level similarly to the repopulation of Andreev levels due to microwave radiation. \cite{bergeret2010,bergeret2011} This process enhances the supercurrent as is shown in Fig. \ref{fig:junction}(e) and (f). The peak in the CPR at phase difference $\varphi_p$ can hence be found from the condition $\omega_L = |\Delta^{+(-)}_{\sigma} (T)|-|\varepsilon_{\sigma}^{+(-)}|$, which can be obtained from Eq. (\ref{eq: AL split}) in the case of $v_S\ll v_0$, and leads to
\begin{equation}
\varphi_p=2\arcsin\bigg\{ \sqrt{\frac{\omega_L}{{\cal D}_0 \Delta(T)}\left( 2-\frac{\omega_L}{\Delta(T)} \right)  }\bigg\} ,
\end{equation}
where $\omega_L\leq {\cal D}_0\Delta(T)/2$. As a consequence, $\varphi_p$ increases as the temperature increases and the superconducting gap, $\Delta(T)$, closes.

\begin{figure}[!htb]
\includegraphics[width=0.97\columnwidth,angle=0]{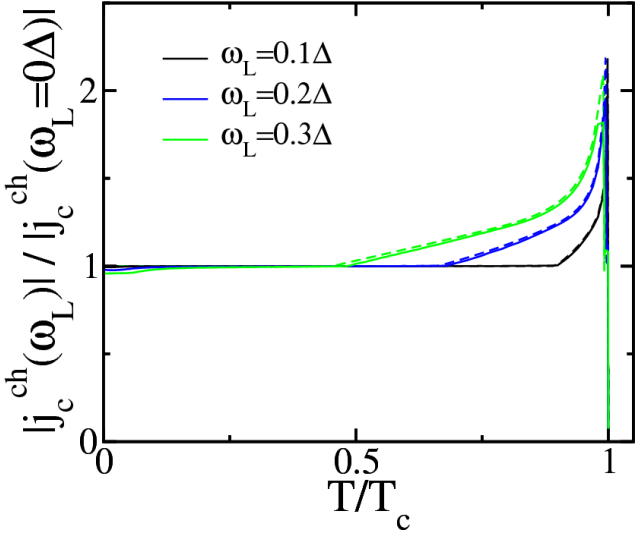}
\caption{\label{fig:IC_T}
(Color online) The spin-precession-assisted charge current in Fig. \ref{fig:junction} may influence the critical charge current, $j_c^{ch}$. This is the case at high temperatures, as shown here where the enhancement of the critical current compared to the critical current for a static spin, $|j_c^{ch}(\omega_L)|/|j_c^{ch}(\omega_L=0\Delta)|$, is plotted as a function of temperature, $T/T_c$, for $(v_0=\cos(0.1\pi/2),v_S=\sin(0.1\pi/2))$ (dashed lines) and $(v_0=0,v_S=1)$ (solid lines). The precession angle is $\vartheta=\pi/8$ and the temperature behavior of the critical current for a static spin is given by $j_c^{ch}(\omega_L=0\Delta) \propto \Delta (T) \tanh \left[ \frac{\Delta(T)}{2T} \right]$. \cite{ambegaokar1963}}
\end{figure}

\begin{figure}[t]
\includegraphics[width=0.97\columnwidth,angle=0]{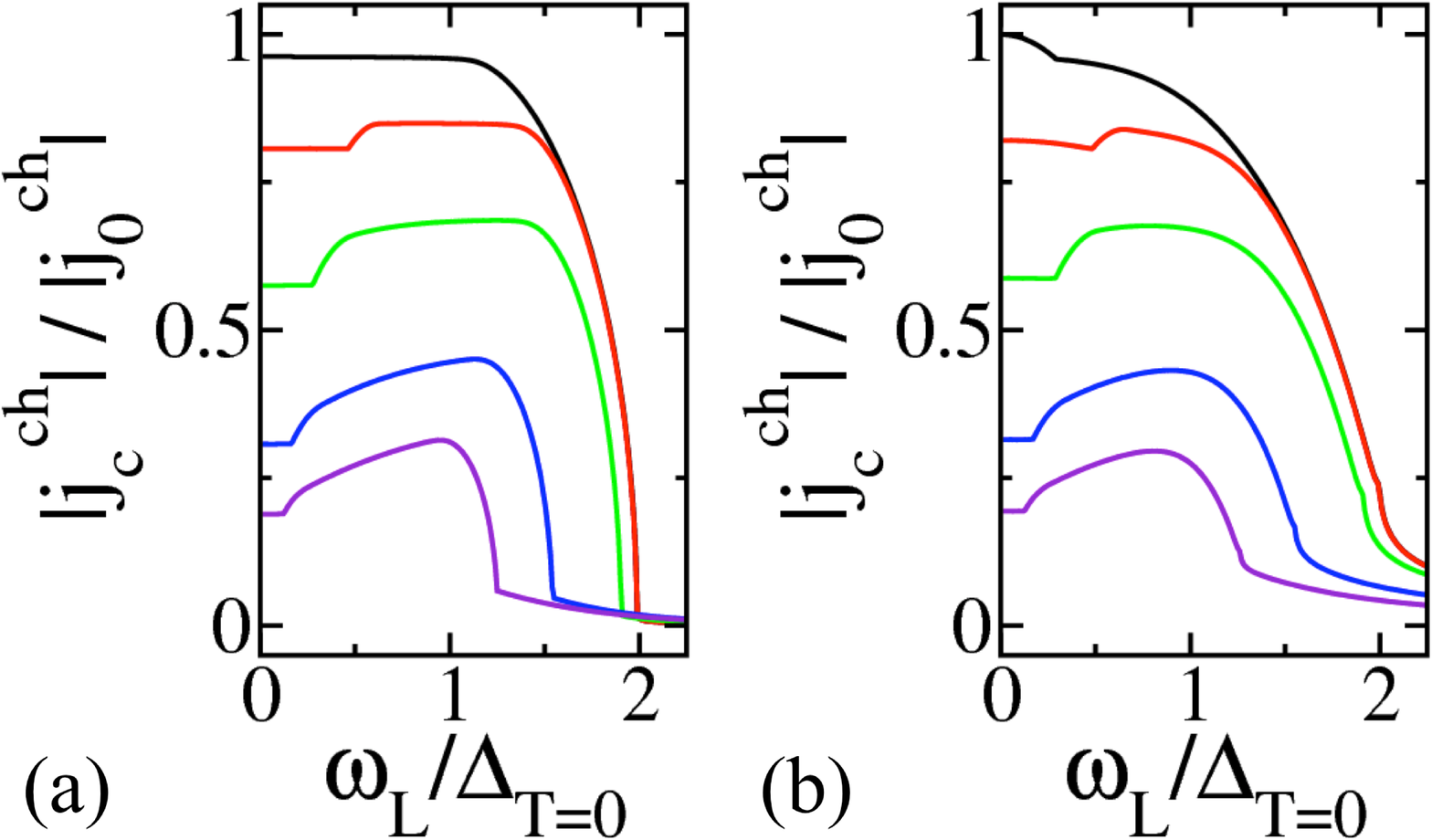}
\caption{\label{fig:IC_omega}
(Color online) Critical current as function of precession frequency, $\omega_L$, for (a) ($v_0=\cos(0.1\pi/2), v_S=\sin(0.1\pi/2)$) and (b) ($v_0=0, v_S=1$), where the precession angle is taken to be $\vartheta=\pi/8$ and the temperature is $T/T_c=7.5\cdot10^{-4}$ (black), $0.25$ (red), $0.50$ (green), $0.75$ (blue) and $0.85$ (violet). The increase of the critical current at low frequencies is an effect of the spin-precession enhancement of the current which occurs at frequencies high enough to couple the lower (upper) Andreev levels with continuum states below (above) the gap edge, i.e. $\omega_L \geq |\Delta^\pm_\sigma(T)|-|\varepsilon^\pm_\sigma(T,\varphi)|$. The decrease of the critical current at frequencies $\omega_L \sim 2 \Delta$ is due to coupling between the lower (upper) Andreev levels with the upper (lower) continuum states.}
\end{figure}

The enhancement of the charge current for certain phase differences may lead to an enhanced critical current if the enhancement is large enough. Fig. \ref{fig:IC_T} shows the enhancement of the critical charge current, $|j_c^{ch}(\omega_L)|/|j_c^{ch}(\omega_L=0)|$, as function of temperature, $T/T_c$, for precession frequencies $\omega_L=0.1\Delta$, $0.2\Delta$ and $0.3\Delta$. Dashed lines denote hopping amplitudes $(v_0=\cos(0.1\pi/2),v_S=\sin(0.1\pi/2))$ while solid lines denote $(v_0=0,v_S=1)$. The enhancement of the critical current may be understood as an effective lowering of the temperature due to the repopulation of the lower Andreev levels and the emptying of the upper Andreev levels - the Andreev level population generated by the precessing spin at a certain temperature, $\phi^{i/o,<}_\sigma(\omega_L>0,T_1)$, corresponds to an Andreev level population at a lower temperature but with zero precession frequency, $\phi^{i/o,<}_\sigma(\omega_L=0,T_2<T_1)$.

The critical charge current as a function of precession frequency is shown in Fig. \ref{fig:IC_omega} for hopping amplitudes (a) ($v_0=\cos(0.1\pi/2), v_S=\sin(0.1\pi/2)$) and (b) ($v_0=0, v_S=1$). The critical current is enhanced at high enough precession frequencies due to the redistribution of the quasiparticle occupation. At even higher precession frequencies, quasiparticles can be scattered between the lower Andreev levels, $\varepsilon^-_{\sigma}$, and the upper continuum states, similarly to what was found in Refs. \onlinecite{bergeret2010} and \onlinecite{bergeret2011}. These processes lead to a decrease of the critical current at precession frequencies $\sim 2\Delta(T)$.

\subsection{Spin currents and spin-triplet correlations}\label{subsec:spin currents and spin-triplet correlations}

In Refs. \onlinecite{holmqvist2011} and \onlinecite{teber2010}, it was found that the spin-dependent scattering across the junction produces a spin structure of the full propagators $\check{g}^{i,o}_\alpha$. In general, the Keldysh-Nambu-spin matrices of the boundary condition problem have a spin structure that enables them be divided into generalized diagonal matrices, $\check{X}^d$, spin-raising matrices, $\check{X}^\uparrow$, and spin-lowering matrices, $\check{X}^\downarrow$. This spin structure of the $\check{g}^{i,o}_\alpha$ matrices, in particular, is what leads to the non-zero spin current of Eq. (\ref{spincurr}). The spin current exists close to the junction interface and decays on the scale of the superconducting coherence length. In Refs. \onlinecite{holmqvist2011} and \onlinecite{teber2010}, the spin current was found to have a precessing polarization and a term due to the spin-dependent Andreev scattering that can be expressed as $j_{\alpha,H}^s(\gamma\Heff)\times\SMM$, which is finite only for temperatures $T<T_c$. This term generates a feed-back effect on the precessing spin in the form of a spin-transfer torque \cite{slonczewski1996,berger1996} given by $\torque_H=(j_{L,H}^s-j_{R,H}^s)(\gamma\Heff)\times\SMM$. \cite{gilbert2004,tserkovnyak2005} This spin-transfer torque acts as an effective magnetic field and shifts the precession frequency as $\omega_L\rightarrow \omega_L[1+2j_H^s]$, where $j_H^s=j_{L,H}^s=-j_{R,H}^s$.

\begin{figure*}[t]
\includegraphics[width=1.90\columnwidth,angle=0]{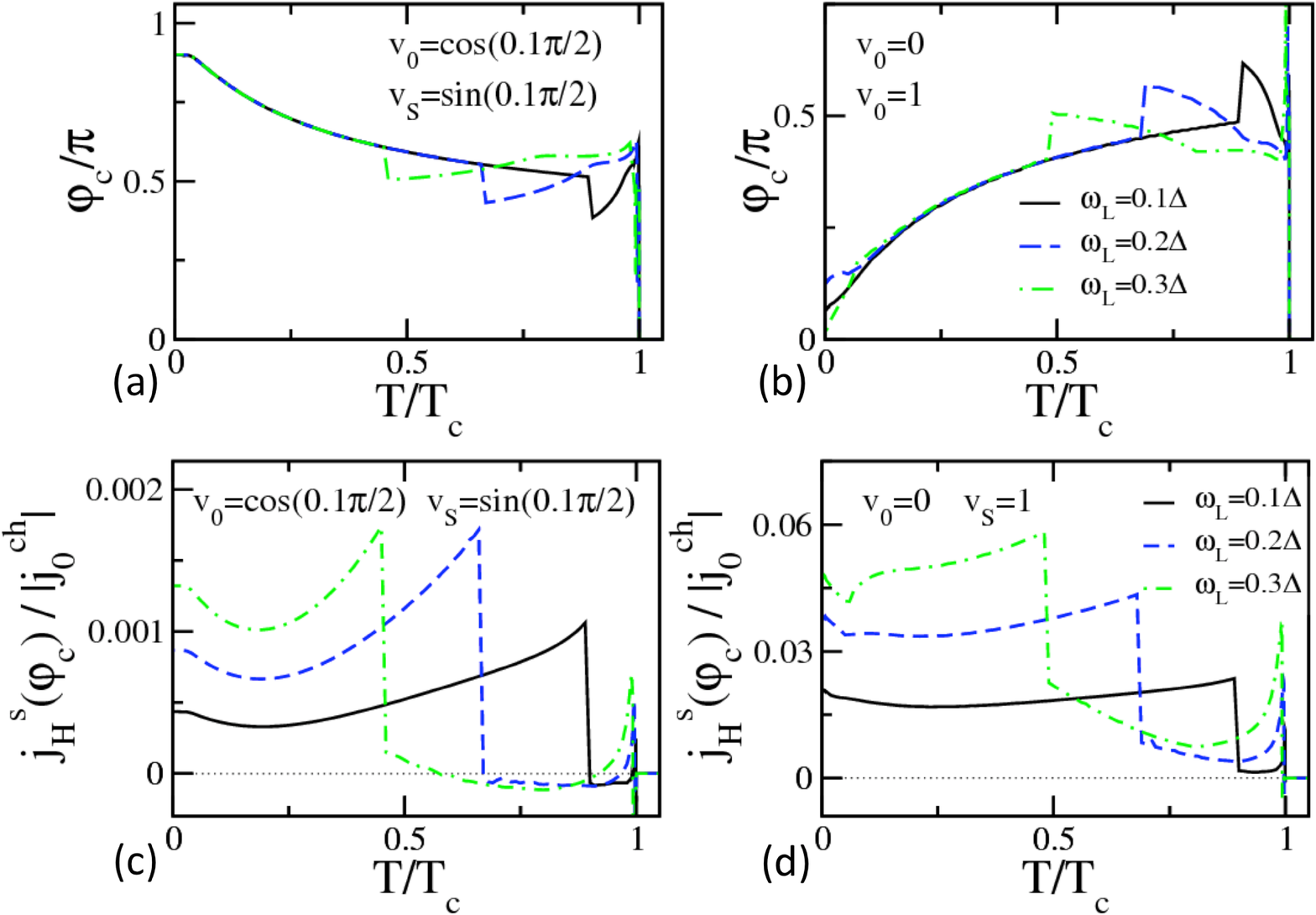}
\caption{\label{fig:d-vec}
(Color online) In panels (a) and (b), the superconducting phase difference, $\varphi_c$, corresponding to the critical charge current, i.e. $j_c^{ch}=j^{ch}(\varphi_c)$, is plotted as a function of temperature, $T/T_c$, for (a) ($v_0=\cos(0.1\pi/2), v_S=\sin(0.1\pi/2)$) and (b) ($v_0=0, v_S=1$) for precession frequencies $\omega_L=0.1\Delta$, $0.2\Delta$ and $0.3\Delta$. The abrupt jumps in the phase difference occurs when the critical current is given by the spin-precession-enhanced current. The abrupt change in $\varphi_c$ leads to an abrupt change in the spin-current component $j_H^s(\varphi_c)$ which is plotted for hopping strengths (c) ($v_0=\cos(0.1\pi/2), v_S=\sin(0.1\pi/2)$) and (d) ($v_0=0, v_S=1$). The spin-current component $j_H^s(\varphi_c)$ modifies the precession frequency $\omega_L$ \cite{holmqvist2011,teber2010} and according to panels (c) and (d), the magnitude of this frequency shift is drastically modified as a function of temperature.
The precession angle is $\vartheta=\pi/8$ in all panels.}
\end{figure*}


The superconducting phase difference $\varphi_c$ corresponding to the critical current, i.e. $j^{ch}_c=j^{ch}(\varphi_c)$, is plotted in Fig. \ref{fig:d-vec} for (a) ($v_0=\cos(0.1\pi/2), v_S=\sin(0.1\pi/2)$) and for (b) ($v_0=0, v_S=1$). As can be seen in the figure, the phase $\varphi_c$ changes abruptly as $\varphi_c$ coincides with $\varphi_p$. The abrupt change in $\varphi_c$ is accompanied with an equally abrupt change in the spin-current component $j_H^s$, which is shown in panels (c) and (d) of Fig. \ref{fig:d-vec} for the two choices of hopping amplitudes.

The connection between the spin currents and the anomalous Green's functions generated by the precessing spin can be described by relatively simple relations in the case of a small tilt angle, $\vartheta = \delta \vartheta$. If the t matrix and the incoming (outgoing) propagator in the case of $\vartheta=0$ are $\check{t}^0_\alpha$ and $\check{g}^{0,i(o)}_\alpha$, respectively, then one can show that (see Appendix \ref{sec: appendix}) a small tilt angle $\delta \vartheta \gtrsim 0$ gives a contribution, $\delta \check{t}$ and $\delta \check{g}$, to first order in $\delta \vartheta$ such that
\begin{eqnarray}
\check{t}_\alpha &=& \check{t}^0_\alpha +\delta \check{t}_\alpha \\ \nonumber
{\rm and}\quad \check{g}^{i/o}_\alpha &=& \check{g}^{0,i/o}_\alpha + \delta \check{g}^{i/o}_\alpha.
\end{eqnarray}
The first-order contribution to the t matrix has the spin structure $\delta \check{t}_\alpha=\delta \check{t}^\uparrow_\alpha + \delta \check{t}^\downarrow_\alpha$ and hence the contribution to the propagators takes the similar form $\delta \check{g}^{i/o}_\alpha=\delta \check{g}^{\uparrow,i/o}_\alpha + \delta \check{g}^{\downarrow,i/o}_\alpha$ where
\begin{eqnarray}\label{eq: delta g updown}
\delta\check{g}^{\uparrow,i/o}_\alpha&=&(\check{g}^0_\alpha\pm i\pi\check{1}) \delta \check{t}^\uparrow_\alpha (\check{g}^0_\alpha \mp i\pi\check{1}) \\ \nonumber
{\rm and\quad} \delta\check{g}^{\downarrow,i/o}_\alpha&=&(\check{g}^0_\alpha\pm i\pi\check{1}) \delta \check{t}^\downarrow_\alpha (\check{g}^0_\alpha \mp i\pi\check{1}).
\end{eqnarray}
Using Eqs. (\ref{eq: g in out}) and (\ref{eq: delta g updown}), one can write the normal and anomalous Green's functions of $\check{g}^{i/o}_\alpha=\check{g}^{0,i/o}_\alpha+\delta \check{g}^{i/o}_\alpha$ as (see Eq. (\ref{scalar and triplet green's function}))
\begin{equation}\label{eq: g 2x2}
g^{i/o,X}_s+\gvec_t^{i/o,X}\cdot\svec=\left(\begin{array}{cc}g^{0,i/o,X}_+ & \delta g^{i/o,X}_\uparrow \\ \delta g^{i/o,X}_\downarrow & g^{0,i/o,X}_-  \end{array}\right)
\end{equation}
and
\begin{equation}\label{eq: f 2x2}
(f^{i/o,X}_s+\fvec_t^{i/o,X}\cdot\svec)i\sy=\left(\begin{array}{cc}-\delta f^{i/o,X}_\uparrow & f^{0,i/o,X}_+ \\ -f^{0,i/o,X}_- & \delta f^{i/o,X}_\downarrow  \end{array}\right).
\end{equation}

The normalization condition,
\begin{equation}\label{norm condition}
\check{g}^2 = -\pi^2 \check{1},
\end{equation}
can be used to relate the anomalous and the normal Green's functions. To first order in $\delta \vartheta$, the normalization condition gives for the retarded (R) and advanced (A) components (of both the incoming and outgoing propagators)
\begin{eqnarray}
\delta g^{R/A}_\uparrow &=& \tilde{F}^{R/A}_- \delta f^{R/A}_\uparrow  + F^{R/A}_+  \delta \tilde{f}^{R/A}_\uparrow \label{gRA up} \\ 
\delta g^{R/A}_\downarrow &=&  \tilde{F}^{R/A}_+ \delta f^{R/A}_\downarrow + F^{R/A}_-  \delta \tilde{f}^{R/A}_\downarrow \label{gRA down}
\end{eqnarray}
where the coefficients $F^{R/A}_{+/-}$ and $\tilde{F}^{R/A}_{\pm}$ are given by Eq. (\ref{eq: FRA}) in the Appendix. The Keldysh components are
\begin{eqnarray}
\delta g^{K}_\uparrow &=& \delta f^R_\uparrow \tilde{F}'^K_- + \delta \tilde{f}^A_\uparrow F'^K_+   - \delta \tilde{f}^R_\uparrow F^R_+ G^K_-  - \delta f^A_\uparrow \tilde{F}^A_- G^K_+ \nonumber \label{gKup}  \\ 
&&   +\delta f^K_\uparrow \tilde{F}'^A_- + \delta \tilde{f}^K_\uparrow F'^R_+    \\ \nonumber
\delta g^{K}_\downarrow &=& \delta f^R_\downarrow \tilde{P}'^K_+ + \delta \tilde{f}^A_\downarrow P'^K_- - \delta \tilde{f}^R_\downarrow F^R_- Q^K_+  - \delta f^A_\downarrow \tilde{F}^A_+ Q^K_-   \\ 
&& \delta f^K_\downarrow \tilde{P}'^A_+ + \delta \tilde{f}^K_\downarrow P'^R_- , \label{gKdown}
\end{eqnarray}
where the retarded-, advanced-type coefficients, $\tilde{F}'^{R/A}_\pm$, $F'^{R/A}_\pm$, $\tilde{P}'^{R/A}_\pm$, $P'^{R/A}_\pm$, and Keldysh-type coefficients, $\tilde{F}'^{K}_\pm$, $F'^{K}_\pm$, $\tilde{P}'^{K}_\pm$, $P'^{K}_\pm$, $G^{K}_\pm$ and $Q^{K}_\pm$, are given by Eqs. (\ref{eq: F'RA}) and (\ref{eq: GK-coeff}) in the Appendix.

The induced spin-triplet components were quantified using $\dvec$ vectors in Ref. \onlinecite{holmqvist2011} as these are used to characterize the order parameters of spin-triplet superconductors. However, the appearance of induced spin-triplet correlations is related to the transport properties of the junction as well. The charge current in Eq. (\ref{chargecurr}) is given by the diagonal elements of $\hat{g}^{i/o,<}_\alpha$. To first order in $\delta \vartheta$, the charge current is then given by
\begin{equation}
j^{ch}_\alpha(t) = \frac{e}{2\hbar} \int \frac{d \varepsilon}{8 \pi i} \mbox{Tr} [ \hat{\tau}_3 
(\hat{g}^{0,i,<}_{\alpha}(\varepsilon,t)  - \hat{g}^{0,o,<}_{\alpha}(\varepsilon,t)) ],
\end{equation}
i.e. the charge current is identical to its value in the $\vartheta=0$ case. The spin current in Eq. (\ref{spincurr}) is given by the triplet-components, which also include the off-diagonal elements of the normal Green's functions, and since the z component of $\delta \check{g}^{i/o}_\alpha$ is zero, the spin-up and -down components of the spin current can explicitly be written as
\begin{equation}\label{spincurr up down}
j^{s}_{\uparrow(\downarrow),\alpha}(t) = \frac{1}{4}  \int \frac{d \varepsilon}{4 \pi i} [  
( \delta g^{i,<}_{\uparrow(\downarrow),\alpha}(\varepsilon,t) - \delta g^{o,<}_{\uparrow(\downarrow),\alpha}(\varepsilon,t)) ],
\end{equation}
where the symmetry relations for the retarded, advanced and Keldysh Green's functions were used \cite{serene1983} and the "$<$" components were obtained as $x^<=\frac{1}{2}(x^K-x^R+x^A)$. From Eqs. (\ref{gRA up}) $-$ (\ref{gKdown}) and (\ref{spincurr up down}), it is clear that the anomalous Green's functions $\check{f}^{\uparrow/\downarrow}$ induced by the precessing spin are what determine the spin current. Conversely, the spin current vanishes, $j^{s}_{\uparrow(\downarrow),\alpha}(t) \equiv 0$, for non-existent anomalous Green's functions, i.e. $\check{f}^{\uparrow/\downarrow} \equiv 0$.

\section{Conclusions}\label{sec: conclusions}

In summary, we have studied the non-equilibrium transport properties of a superconducting point contact coupled to the precessing spin of a nanomagnet. First, we analyzed the Andreev levels and their population as a function of temperature. It was found that the precession of the nanomagnet modifies the Andreev scattering in such a way that it leads to scattering of quasiparticles from below the superconducting gap edge into the lower Andreev levels. These transitions are similar those occurring in microwave-irrated superconducting quantum point contacts \cite{bergeret2010,bergeret2011} and lead to a non-equilibrium population of the Andreev levels and an effective cooling of the point contact. 

We have also shown that the effective cooling leads to an enhanced supercurrent. If the precession frequency of the nanomagnet's spin is large enough, this spin-precession-assisted supercurrent has an enhanced critical current. The enhancement of the critical current due to the spin precession increases with increasing temperature.

The scattering across the junction generated by the precessing spin leads to additional non-equilibrium processes: Besides the enhanced supercurrent, we have also shown that spin-triplet correlations are created as a response to the scattering processes generated by the spin precessing at a finite angle. Moreover, we have shown that the induced spin-triplet correlations create a spin current. As was shown previously, \cite{holmqvist2011,teber2010} one component of the spin current shifts the precession frequency of the spin. Here, this spin-current component was studied for the same superconducting phase difference that is associated with the critical current and the magnitude of this component was found to be drastically modified as a function of temperature.
The critical current enhancement is an effect of the non-equilibrium processes taking place inside the junction and a measurement of this enhancement would suggest the existence of induced spin-triplet correlations since both phenomena have the same origin.

A natural question concerns the experimental control over the junction parameters. Aluminum microbridges can be used to fabricate few-channel superconducting atomic point contacts \cite{goffman2000} whose transmission eigenvalues are possible to determine via transport measurements. \cite{scheer1997,scheer2001} The superconducting gap in aluminum is $\sim 200\,\mu$eV but can be made smaller in an atomic point contact.
Typical values for precession frequencies are in the range of tenths of GHz for ferromagnetic-resonance experiments performed on thin ferromagnetic films in contact with superconductors. \cite{bell2008} With an applied external magnetic field of $\sim 60$ mT, the corresponding gyromagnetic ratio is close to that of free electrons, $\gamma=2\mu_B/\hbar$. Assuming similar values for the nanomagnet, the precession frequency is in the range $\omega_L/ \Delta\sim 0.07-0.3$ for a junction with a superconducting gap in the range $\Delta \sim 20\,\mu$eV$-100\,\mu$eV.

A direct measurement of the current-phase relation can in principle be done for atomic point contacts \cite{dellarocca2007}. However, this type of measurement requires the addition of a SQUID loop to control the phase difference of the superconducting junction and the magnetic flux through the loop might interfere with the magnetic control of the dynamics of the nanomagnet.
Therefore, measurements of the critical current as a function of temperature should be the more practical route to finding experimental signatures of the coupling between a superconducting junction and the dynamics of a nanomagnet and could lead to more insight into the interplay between superconductivity and ferromagnetism.

\section*{Acknowledgments}
C.~H. and W.~B. were supported by Deutsche Forschungsgemeinschaft and SFB 767. M.~F. acknowledges support from the Swedish Research Council (VR).

\appendix
\section{Calculation of spin-triplet correlations}\label{sec: appendix}

In the limit of a small precessing angle, $\vartheta = \delta \vartheta$, the contributions to the t matrices and propagators can be calculated as follows:
Starting with a junction with zero tilt angle, $\vartheta=0$, the hopping element takes the diagonal form
\begin{equation}\label{hopping vd}
\hat{v}=\left(\begin{array}{cc}v_0+v_S \sz &0 \\ 0 & v_0+v_S  \sz \end{array}\right)\equiv\hat{v}^d
\end{equation}
and the corresponding t-matrix equation is
\begin{equation}\label{t-matrix t0}
\check{t}^0_\alpha(\varepsilon)=\check{\Gamma}^0_\alpha (\varepsilon)+[ \check{\Gamma}^0_\alpha \circ \check{g}^0_\alpha \circ \check{t}^0_\alpha ] (\varepsilon)
\end{equation}
where $\alpha$ denotes the left (L) or right (R) lead, $\check{g}^0_\alpha$ is the unperturbed Green's function and $\check{\Gamma}^0_{L/R}=\check{v}^d\circ \check{g}^0_{R/L} \circ \check{v}^d$. This equation has the solution $\check{t}^0_\alpha=[\check{1}- \check{\Gamma}^0_\alpha  \check{g}^0_\alpha]^{-1} \check{\Gamma}^0_\alpha$. Notice that the t matrix $\check{t}^0_\alpha$ has a diagonal form due to the spin structure of $\check{\Gamma}^0_\alpha$. The incoming and outgoing propagators, $\check{g}^i$ and $\check{g}^o$, given by
\begin{equation}\label{g0}
\check{g}^{0,i/o}_\alpha=\check{g}^0_\alpha+(\check{g}^0_\alpha\pm i\pi\check{1}) \check{t}^0_\alpha (\check{g}^0_\alpha \mp i\pi\check{1}),
\end{equation}
have the same diagonal form as $\check{t}^0_\alpha$ and yield a spin current $\jvec^s_\alpha=0$ (see Eq. (\ref{spincurr})).

If a small tilt angle, $\vartheta=\delta \vartheta$, is introduced, the hopping element is modified into
\begin{equation}
\hat{v}=\left(\begin{array}{cc}v_0+v_S \sz + v_S \delta\vartheta \sx &0 \\ 0 & v_0+v_S \sz  + v_S \delta\vartheta \sx \end{array}\right),
\end{equation}
where it was used that $\cos(\delta \vartheta)\approx 1$ and $\sin(\delta \vartheta)\approx\delta\vartheta$. The hopping element, $\hat{v}$, can hence be divided into $\hat{v}=\hat{v}^d+\hat{v}^\uparrow+\hat{v}^\downarrow$, where $\hat{v}^d$ is given by Eq. (\ref{hopping vd}) and both $\hat{v}^\uparrow$ and $\hat{v}^\downarrow$ are proportional to $\delta \vartheta$:
\begin{equation}
\hat{v}^{\uparrow(\downarrow)}=\left(\begin{array}{cc}v_S \delta\vartheta \sigma_{+(-)} &0 \\ 0 &  v_S \delta\vartheta \sigma_{-(+)} \end{array}\right).
\end{equation}
With this hopping element, the change in the t matrix can be written as 
\begin{equation}\label{t+dt}
\check{t}_\alpha=\check{t}^0_\alpha+\delta \check{t}_\alpha,
\end{equation}
where $\check{t}^0_\alpha$ is the solution of Eq. (\ref{t-matrix t0}) and $\delta \check{t}_\alpha$ is due to the tilt angle, $\delta\vartheta$. Similarly,
\begin{equation}\label{G+dG}
\check{\Gamma}_\alpha=\check{\Gamma}^0_\alpha+ \delta \check{\Gamma}_\alpha,
\end{equation}
where $\check{\Gamma}^0_{L/R}= \check{v}^d \check{g}_{R/L}^0 \check{v}^d$ and the tilt angle gives the contribution $\delta \check{\Gamma}_\alpha$. However, the diagonal component of $\delta \check{\Gamma}_\alpha$ is $\delta \check{\Gamma}^d_{L/R}=\hat{v}^\uparrow \check{g}_{R/L}^0 \hat{v}^\downarrow + \hat{v}^\downarrow \check{g}_{R/L}^0 \hat{v}^\uparrow\approx 0$ to first order in $\delta \vartheta$. Then, $\delta \check{\Gamma}_\alpha = \delta \check{\Gamma}^\uparrow + \delta \check{\Gamma}^\downarrow$ where $\delta \check{\Gamma}^\uparrow_{L/R}=\hat{v}^\uparrow \check{g}_{R/L}^0 \hat{v}^d+\hat{v}^d \check{g}_{R/L}^0 \hat{v}^\uparrow$ and $\delta \check{\Gamma}^\downarrow_{L/R}=\hat{v}^\downarrow \check{g}_{R/L}^0 \hat{v}^d+\hat{v}^d \check{g}_{R/L}^0 \hat{v}^\downarrow$, which means that $\delta \check{\Gamma}^{\uparrow(\downarrow)}_\alpha \propto \delta\vartheta$. Inserting Eqs. (\ref{t+dt}) and (\ref{G+dG}) into the t-matrix equation and subtracting Eq. (\ref{t-matrix t0}) gives
\begin{equation}
\delta \check{t}_\alpha=\delta \check{\Gamma}_\alpha [\check{1}+\check{g}^0_\alpha \check{t}^0_\alpha] + \check{\Gamma}^0_\alpha \check{g}^0_\alpha \delta \check{t}_\alpha
\end{equation}
to first order in $\delta\vartheta$. Due to the spin structure of $\delta \check{\Gamma}_\alpha$, the t matrix is $\delta \check{t}_\alpha=\delta \check{t}^\uparrow_\alpha+\delta \check{t}^\downarrow_\alpha$ where
\begin{equation}
\delta \check{t}^{\uparrow(\downarrow)}_\alpha=[\check{1}-\check{\Gamma}^0_\alpha \check{g}^0_\alpha]^{-1} \delta \check{\Gamma}^{\uparrow(\downarrow)} [\check{1}+\check{g}^0_\alpha \check{t}^0_\alpha].
\end{equation}
The incoming and outgoing propagators are then given by $\check{g}^{i/o}_\alpha=\check{g}^{0,i/o}_\alpha+\delta \check{g}^{i/o}_\alpha$, where $\delta \check{g}^{i/o}_\alpha=\delta \check{g}^{\uparrow,i/o}_\alpha+\delta \check{g}^{\downarrow,i/o}_\alpha$ and
\begin{eqnarray}\label{delta g}
\delta\check{g}^{\uparrow,i/o}_\alpha&=&(\check{g}^0_\alpha\pm i\pi\check{1}) \delta \check{t}^\uparrow_\alpha (\check{g}^0_\alpha \mp i\pi\check{1}) \\ \nonumber
{\rm and\quad} \delta\check{g}^{\downarrow,i/o}_\alpha&=&(\check{g}^0_\alpha\pm i\pi\check{1}) \delta \check{t}^\downarrow_\alpha (\check{g}^0_\alpha \mp i\pi\check{1}).
\end{eqnarray}

Using the normalization condition, Eq. (\ref{norm condition}), and the explicit forms of the normal and anomalous Green's functions given by Eqs. (\ref{eq: g 2x2}) and (\ref{eq: f 2x2}), respectively, the retarded and advanced normal Green's functions can be expressed in terms of the anomalous ones as
\begin{eqnarray}
\delta g^{R/A}_\uparrow &=& \tilde{F}^{R/A}_- \delta f^{R/A}_\uparrow  + F^{R/A}_+  \delta \tilde{f}^{R/A}_\uparrow  \\ 
\delta g^{R/A}_\downarrow &=&  \tilde{F}^{R/A}_+ \delta f^{R/A}_\downarrow + F^{R/A}_-  \delta \tilde{f}^{R/A}_\downarrow 
\end{eqnarray}
where
\begin{eqnarray}\label{eq: FRA}
F^{R/A}_{\pm}&=&f^{0,R/A}_{\pm}/\xi^{0,R/A} \\ \nonumber
\tilde{F}^{R/A}_{\pm}&=&\tilde{f}^{0,R/A}_{\pm}/\xi^{0,R/A}.
\end{eqnarray}
Here, $g^X_{\pm}$ and $f^X_{\pm}$ are the diagonal matrix components given by $g^X_{\pm}=g^X_{s}\pm g^X_{z}$ and $f^X_{\pm}=f^X_{s}\pm f^X_{z}$ and $\xi^{R/A}=g^{0,R/A}_+ + g^{0,R/A}_-$.
The off-diagonal Keldysh components are
\begin{eqnarray}
\delta g^{K}_\uparrow &=& \delta f^R_\uparrow \tilde{F}'^K_- + \delta \tilde{f}^A_\uparrow F'^K_+   - \delta \tilde{f}^R_\uparrow F^R_+ G^K_-  - \delta f^A_\uparrow \tilde{F}^A_- G^K_+ \nonumber   \\ 
&&   +\delta f^K_\uparrow \tilde{F}'^A_- + \delta \tilde{f}^K_\uparrow F'^R_+    \\ \nonumber
\delta g^{K}_\downarrow &=& \delta f^R_\downarrow \tilde{P}'^K_+ + \delta \tilde{f}^A_\downarrow P'^K_- - \delta \tilde{f}^R_\downarrow F^R_- Q^K_+  - \delta f^A_\downarrow \tilde{F}^A_+ Q^K_-   \\ 
&& \delta f^K_\downarrow \tilde{P}'^A_+ + \delta \tilde{f}^K_\downarrow P'^R_- 
\end{eqnarray}
where the retarded-type and advanced-type matrices are
\begin{eqnarray}\label{eq: F'RA}
F'^{R/A}_{\pm}=f^{0,R/A}_{\pm}/\zeta_1 ; &\,&  P'^{R/A}_\pm=f^{0,R/A}_\pm /\zeta_2  \\ \nonumber
 \\ \nonumber
\tilde{F}'^{R/A}_{\pm}=\tilde{f}^{0,R/A}_{\pm}/\zeta_1  ; &\,& \tilde{P}'^{R/A}_\pm=\tilde{f}^{0,R/A}_\pm /\zeta_2 
\end{eqnarray}
with $\zeta_1=g^{0,R}_+ + g^{0,A}_-$ and $\zeta_2$ is obtained from $\zeta_1$ by exchanging $- \leftrightarrow +$.
The Keldysh-type matrices are
\begin{eqnarray}\label{eq: GK-coeff}
G^K_{\pm}=\zeta_\pm \phi^0_{\pm} /\zeta_1 ; &\,&  Q^K_{\pm}=\zeta_\pm \phi^0_{\pm} /\zeta_2 \\ \nonumber
F'^K_\pm=[ F'^R_\pm - F^A_\pm ]\phi^0_\pm  ; &\,& P'^K_\pm=[ P'^R_\pm - F^A_\pm ]\phi^0_\pm  \\ \nonumber
\tilde{F}'^K_\pm=[ \tilde{F}^R_\pm - \tilde{F}'^A_\pm ]\phi^0_\pm ; &\,& \tilde{P}'^K_\pm=[ \tilde{F}^R_\pm - \tilde{P}'^A_\pm ]\phi^0_\pm \end{eqnarray}
where $\phi^0_\pm$ is the occupation associated with $g^0_\pm$ and $\zeta_{\pm}=g^{0,R}_\pm -g^{0,A}_\pm$.

We have thus shown that a small tilt angle, $\delta \vartheta$, produces spin-flip scattering that combined with Andreev scattering processes leads to spinful anomalous Green's functions, $\delta f^{\uparrow/ \downarrow}$, that in turn generate off-diagonal, spinful normal Green's functions, $\delta g^{\uparrow/ \downarrow}$. Consequently, the spin current given by Eq. (\ref{spincurr}) is non-zero as a result of the induced spin-triplet anomalous Green's functions.


\begin{thebibliography}{9}

\bibitem{baibich1988} M. N. Baibich, J. M. Broto, A. Fert, F. Nguyen Van Dau, F. Petroff, P. Etienne, G. Creuzet, A. Friederich, and J. Chazelas, Phys. Rev. Lett. {\bf 61}, 2472 (1988).

\bibitem{binasch1989} G. Binasch, P. Gr\"unberg, F. Saurenbach, and W. Zinn, Phys. Rev. B {\bf 39}, 4828 (1989).





\bibitem{wolf2001} S.~A. Wolf, D. D. Awschalom, R. A. Buhrman, J. M. Daughton, S. von Moln\'ar, M. L. Roukes, A. Y. Chtchelkanova and D. M. Trege, Science \textbf{294}, 1488 (2001).

\bibitem{elbing2005} M. Elbing, R. Ochs, M. Koentopp, M. Fischer, C. von H\"anisch, F. Weigend, F. Evers, H. B. Weber, and M. Mayor, PNAS \textbf{102}, 8815 (2005).

\bibitem{park2002} J. Park, A. N. Pasupathy, J. I. Goldsmith, C. Chang, Y. Yaish, J. R. Petta, M. Rinkoski, J. P. Sethna, H. D. Abru\~na, P. L. McEuen, and D. C. Ralph, Nature \textbf{417}, 722 (2002).

\bibitem{osorio2008} E. Osorio, T. Bj\o rnholm, J.-M. Lehn, M. Ruben, and H. S. J. van der Zant, J. Phys. Condens. Matter \textbf{20}, 374121 (2008).

\bibitem{yu2005} L. H. Yu, Z. K. Keane, J. W. Ciszek, L. Cheng, J. M. Tour, T. Baruah, M. R. Pederson, and D. Natelson, Phys. Rev. Lett. \textbf{95}, 256803 (2005).

\bibitem{bogani2008}L. Bogani and W. Wernsdorfer, Nat. Mat. {\bf 7}, 179 (2008).

\bibitem{christou2000}G. Christou, D. Gatteschi, D.~N.~Hendrickson, and R.~Sessoli, MRS Bulletin {\bf 25}, 66 (2000).

\bibitem{wernsdorfer1999} W. Wernsdorfer and R. Sessoli, Science {\bf 284}, 133 (1999).

\bibitem{heersche2006} H.~B. Heersche, Z. de Groot, J. A. Folk, H.~S.~J. van der Zant, C. Romeike, M.~R. Wegewijs, L. Zobbi, D. Barreca, E. Tondello, and A. Cornia, Phys. Rev. Lett. {\bf 96}, 206801 (2006).

\bibitem{jo2006} M.-H. Jo, J. E. Grose, K. Baheti, M. M. Deshmukh, J. J. Sokol, E. M. Rumberger, D. N. Hendrickson, J. R. Long, H. Park, and D. C. Ralph, Nano Lett. {\bf 6}, 2014 (2006).

\bibitem{henderson2007} J. J. Henderson, C. M. Ramsey, E. del Barco, A. Mishra, and G. Christou, J. Appl. Phys. {\bf 101}, 09E102 (2007).

\bibitem{grose2008} J. E. Grose, E. S. Tam, C. Timm, M. Scheloske, B. Ulgut, J. J. Parks, H. D. Abru\~na, W. Harneit, and D. C. Ralph, Nature Mater. {\bf 7}, 884 (2008).

\bibitem{zyazin2010} A. S. Zyazin, J. W. G. van den Berg, E. A. Osorio, H. S. J. van der Zant, N. P. Konstantinidis, M. Leijnse, M. R. Wegewijs, F. May, W. Hofstetter, C. Danieli, and A. Cornia, Nano Lett. {\bf 10}, 3307 (2010).

\bibitem{roch2011} N. Roch, R. Vincent, F. Elste, W. Harneit, W. Wernsdorfer, C. Timm, and F. Balestro, Phys. Rev. B {\bf 83}, 081407(R) (2011).

\bibitem{haque2011} F. Haque, M. Langhirt, E. del Barco, T. Taguchi, and G. Christou, J. Appl. Phys. {\bf 109}, 07B112 (2011).

\bibitem{urdampilleta2011} M. Urdampilleta, S. Klyatskaya, J.-P. Cleuziou, M. Ruben, and W. Wernsdorfer, Nat. Mat. {\bf 6}, 185 (2011).

\bibitem{kahle2012}
S. Kahle, Z. Deng, N. Malinowski, C. Tonnoir, A. Forment-Aliaga, N. Thontasen, G. Rinke, D. Le, V. Turkowski, T. S. Rahman, S. Rauschenbach, M. Ternes, and K. Kern, Nano Lett. \textbf{12}, 518 (2012).


\bibitem{eschrig2011} M.~Eschrig, Physics Today {\bf 64}, 43 (2011). 

\bibitem{bulaevskii1977} L. N. Bulaevskii, V. V. Kuzii, and A. A. Sobyanin, PisÕma Zh. Eksp. Teor. Fiz. {\bf 25}, 314 (1977) [JETP Lett. {\bf 25}, 290 (1977)].

\bibitem{ryazanov2001} V. V. Ryazanov, V. A. Oboznov, A. Yu. Rusanov, A. V. Veretennikov, A. A. Golubov, and J. Aarts, Phys. Rev. Lett. {\ 86}, 2427 (2001).

\bibitem{kontos2002} T. Kontos, M. Aprili, J. Lesueur, F. Gen\^et, B. Stephanidis, and R. Boursier , Phys. Rev. Lett. {\bf 89}, 137007 (2002).

\bibitem{buzdin2005} A.~I. Buzdin, Rev. Mod. Phys. \textbf{77}, 935 (2005).

\bibitem{bergeret2001} F.~S. Bergeret, A.~F. Volkov, and K.~B. Efetov, Phys. Rev. Lett. \textbf{86}, 3140 (2001).

\bibitem{bergeret2005} F.~S. Bergeret, A.~F. Volkov, and K.~B. Efetov, Rev. Mod. Phys. \textbf{77}, 1321 (2005).

\bibitem{houzet2007} M. Houzet and A. I. Buzdin, Phys. Rev. B {\bf 76}, 060504(R) (2007).

\bibitem{braude2007} V. Braude and Yu.~V. Nazarov, Phys. Rev. Lett. \textbf{98}, 077003 (2007).

\bibitem{eschrig2008} M. Eschrig and T. L\"ofwander, Nature Physics {\bf 4}, 138 (2008).

\bibitem{keizer2006} R. S. Keizer, S. T. B. Goennenwein, T. M. Klapwijk, G. Miao, G. Xiao, and A. Gupta, Nature {\bf 439}, 825 (2006).

\bibitem{khaire2010} T. S. Khaire, M. A. Khasawneh, W. P. Pratt, Jr., and N. O. Birge, Phys. Rev. Lett. {\bf 104}, 137002 (2010).

\bibitem{waintal2001} X. Waintal and P. W. Brouwer, Phys. Rev. B {\bf 63}, R220407 (2001). 

\bibitem{waintal2002} X. Waintal and P. W. Brouwer, Phys. Rev. B {\bf 65}, 054407 (2002).

\bibitem{zhao2008} E. Zhao and J. A. Sauls, Phys. Rev. B {\bf 78}, 174511 (2008). 

\bibitem{shomali2011} Z. Shomali, M. Zareyan, and W. Belzig, New J. Phys. {\bf 13}, 083033 (2011).


\bibitem{braude2008} V. Braude and Ya.~M. Blanter, Phys. Rev. Lett. \textbf{100}, 207001 (2008).

\bibitem{michelsen2008} J. Michelsen, V. S. Shumeiko, and G. Wendin, Phys. Rev. B {\bf 77}, 184506 (2008).

\bibitem{petkovic2009} I. Petkovi\'c, M. Aprili, S. E. Barnes, F. Beuneu, and S. Maekawa, Phys. Rev. B {\bf 80}, 220502 (2009).

\bibitem{barnes2011} S E Barnes, M Aprili, I Petkovi\'c, S Maekawa, Superconductor Science and Technology {\bf 24}, 024020 (2011).

\bibitem{mai2011} S. Mai, E. Kandelaki, A. F. Volkov, and K. B. Efetov, Phys. Rev. B {\bf 84}, 144519 (2011).

\bibitem{linder2011} J. Linder and T. Yokoyama, Phys. Rev. B {\bf 83}, 012501 (2011).


\bibitem{holmqvist2011} C. Holmqvist, S. Teber, and M. Fogelstr\"om, Phys. Rev. B {\bf 83}, 104521 (2011).

\bibitem{teber2010} S. Teber, C. Holmqvist, and M. Fogelstr\"om, Phys. Rev. B {\bf 81}, 174503 (2010).


\bibitem{kazumov2005} A. Yu. Kasumov, K. Tsukagoshi, M. Kawamura, T. Kobayashi, Y. Aoyagi, K. Senba, T. Kodama, H. Nishikawa, I. Ikemoto, K. Kikuchi, V. T. Volkov, Yu. A. Kasumov, R. Deblock, S. Gu\'eron, and H. Bouchiat, Phys. Rev. B {\bf 72}, 033414 (2005).

\bibitem{houzet2008} M. Houzet, Phys. Rev. Lett. {\bf 101}, 057009 (2008).

\bibitem{bell2008} C.~Bell, S.~Milikisyants, M.~Huber, and J.~Aarts, Phys. Rev. Lett. {\bf 100}, 047002 (2008).

\bibitem{zhu2003} J.~X.~Zhu and A.~V.~Balatsky, Phys. Rev. B {\bf 67}, 174505 (2003).
\bibitem{zhu2004} J.~X.~Zhu, Z. Nussinov, A. Shnirman, and A. V. Balatsky, Phys. Rev. Lett. {\bf 92}, 107001 (2004).


\bibitem{eilenberger1968} G. Eilenberger, Z. Phys. {\bf214}, 195 (1968).

\bibitem{larkin1968} A.~I. Larkin and Y.~N. Ovchinnikov, Zh. \'Eksp. Teor. Fiz. {\bf55}, 2262 (1968).

\bibitem{eliashberg1971} G.~M. Eliashberg, Zh. \'Eksp. Teor. Fiz. {\bf61}, 1254 (1971).



\bibitem{cuevas2001} J.~C. Cuevas, M. Fogelstr\"om, Phys. Rev. B {\bf64}, 104502 (2001).
\bibitem{kopu2004} J. Kopu, M. Eschrig, J. C. Cuevas, and M. Fogelstr\"om, Phys. Rev. B {\bf69}, 094501 (2004).

\bibitem{serene1983} J.~W. Serene and D. Rainer, Phys. Rep. {\bf 101}, 21 (1983).

\bibitem{caroli1971} C. Caroli, R. Combescot, P. Nozieres, and D. Saint-James, J. Phys. C {\bf 4}, 916 (1971).

\bibitem{buchholtz1979} L. J. Buchholtz and D. Rainer, Z. Phys. B 35, 151 (1979).

\bibitem{martinrodero1994} A. Mart\'in-Rodero, F.J. Garc\'ia-Vidal, and A. Levy Yeyati, Phys. Rev. Lett. {\bf 72}, 554 (1994); A. Levy Yeyati, A. Mart\'in-Rodero, and F. J. Garc\'ia-Vidal, Phys. Rev. B {\bf 51}, 3743 (1995).

\bibitem{pillet2010} J.-D. Pillet, C. H. L. Quay, P. Morfin, C. Bena, A. Levy Yeyati, and P. Joyez, Nat. Phys. \textbf{6}, 965 (2010).



\bibitem{shumeiko1997} V. S. Shumeiko, E. N. Bratus, and G. Wendin, Low Temp. Phys. {\bf 23}, 181 (1997).

\bibitem{beenakker1991} C. W. J. Beenakker, Phys. Rev. Lett. {\bf 67}, 3836 (1991).


\bibitem{bergeret2010} F.~S. Bergeret, P. Virtanen, T.~T. Heikkil\"a, and J.~C. Cuevas, Phys.~Rev.~Lett. {\bf105}, 117001 (2010).

\bibitem{bergeret2011} F.~S. Bergeret, P. Virtanen, A. Ozaeta, T.~T. Heikkil\"a, and J.~C. Cuevas, Phys.~Rev.~B {\bf 84}, 054504 (2011).


\bibitem{slonczewski1996} J. C. Slonczewski, J. Magn. Magn. Mater. {\bf 159}, L1 (1996).

\bibitem{berger1996} L. Berger, Phys. Rev. B {\bf 54}, 9353 (1996).

\bibitem{gilbert2004} T. Gilbert, IEEE Trans. Magn. {\bf 40}, 3443 (2004).

\bibitem{tserkovnyak2005} Y. Tserkovnyak, A. Brataas, and G. E. W. Bauer, Phys. Rev. Lett. {\bf 88}, 117601 (2002); Y. Tserkovnyak, A. Brataas, G. E. W. Bauer, and B. I. Halperin, Rev. Mod. Phys. {\bf 77}, 1375 (2005).


\bibitem{ambegaokar1963} V.~Ambegaokar and A.~Baratoff, Phys. Rev. Lett. {\bf 10}, 486 (1963); erratum, {\bf 11}, 104 (1963).


\bibitem{goffman2000} M. F. Goffman, R. Cron, A. Levy Yeyati, P. Joyez, M. H. Devoret, D. Esteve, and C. Urbina, Phys. Rev. Lett. {\bf 85}, 170 (2000).

\bibitem{scheer1997} E. Scheer, P. Joyez, D. Esteve, C. Urbina, and M. H. Devoret, Phys. Rev. Lett. {\bf 78}, 3535 (1997). 

\bibitem{scheer2001}  E. Scheer, W. Belzig, Y. Naveh, M.~H. Devoret, D. Esteve, and C. Urbina,
Phys. Rev. Lett. \textbf{86}, 284 (2001).


\bibitem{dellarocca2007} M. L. Della Rocca, M. Chauvin, B. Huard, H. Pothier, D. Esteve, C. Urbina, Phys. Rev. Lett. {\bf 99}, 127005 (2007).



\end{thebibliography}
\end{document}